\documentclass[12pt]{iopart}
\usepackage{amssymb,amsmath}
\usepackage{color} 
\usepackage{accents} 
\usepackage{texdraw}
\usepackage{eucal}
\usepackage{epic,epsfig}
\usepackage{tikz}
\usetikzlibrary{positioning}

\DeclareMathAccent{\wtilde}{\mathord}{largesymbols}{"65}
\DeclareMathAccent{\what}{\mathord}{largesymbols}{"62}

\def\m@th{\mathsurround=0pt}
\mathchardef\bracell="0365 
\def\upbrall{$\m@th\bracell$}
\def\undertilde#1{\mathop{\vtop{\ialign{##\crcr
    $\hfil\displaystyle{#1}\hfil$\crcr
     \noalign
     {\kern1.5pt\nointerlineskip}
     \upbrall\crcr\noalign{\kern1pt
   }}}}\limits}
\def\theequation{\arabic{section}.\arabic{equation}}

\newcommand{\bde}{\boldsymbol{e}}

\newcommand{\bn}{\boldsymbol{n}}

\newcommand{\al}{\alpha}

\newcommand{\dl}{\delta}
\newcommand{\Dl}{\Delta}

\newcommand{\sg}{\sigma}

\newcommand{\cL}{\mathcal{L}}

\newcommand{\ptl}{\partial}

\newcommand{\Li}{{\rm Li}}

\newcommand{\bblu}{\begin{color}{blue}}
\newcommand{\bred}{\begin{color}{red}}
\newcommand{\ecl}{\end{color}}

\newcommand{\nn}{\nonumber}

\newcommand{\be}{\begin{equation}}
\newcommand{\ee}{\end{equation}}
\newcommand{\bea}{\begin{eqnarray}}
\newcommand{\eea}{\end{eqnarray}}
\newcommand{\bse}{\begin{subequations}}
\newcommand{\ese}{\end{subequations}}

\begin{document}
\title{Lagrangian multiforms and multidimensional consistency} 
\author{Sarah Lobb and Frank Nijhoff}
\address{Department of Applied Mathematics, University of Leeds, Leeds LS2 9JT, UK}

\begin{abstract}
We show that well-chosen Lagrangians for a class of two-dimensional integrable lattice equations obey a closure relation when embedded in a higher dimensional lattice. On the basis of this property we formulate a Lagrangian description for such systems in terms of Lagrangian multiforms. We discuss the connection of this formalism with the notion of multidimensional consistency, and the role of the lattice 
 from the point of view of the relevant variational principle.
\end{abstract}

\section{Introduction}
In recent years there has been a growing interest in the integrability of discrete systems defined on two- or multidimensional lattices. In part the study of such systems may be motivated by the search for accurate approximations to continuous systems. However, the modern point of view is that such lattice systems are important in their own right from a theoretical perspective, and, in fact, are thought to be richer and more generic than their continuous counterparts. Discrete systems have also been proposed in physics to model the fundamental interactions on  the scale of the Planck constant where space and time themselves can be thought of as being discrete \cite{Lee,'tHooft}.

The earliest examples of integrable lattice systems go back to the mid 1970s and early 1980s, when the research was focused on discretizing known continuous soliton systems \cite{Abl+L1,Abl+L2,Hirota1,Hirota2,DJM,Homotopy,QNCL}. In recent years, the insight has developed that the key aspect of integrability resides in the property of multidimensional consistency \cite{CAC,B+S,Frank}. This property entails that an equation can be embedded in a consistent way in a multidimensional lattice, i.e. by imposing a copy of the equation (with appropriate \emph{lattice parameters}) in each pair of directions, such that there is no inconsistency or multivaluedness occurring in the evaluation of the dependent variables on each lattice site.
Using this property a classification of two-dimensional scalar integrable lattice systems was given, in the affine linear case, by Adler, Bobenko and Suris \cite{ABS,ABS2} (resulting in what is hereafter referred to as the ABS list). In addition to the known examples of lattice systems of KdV type, cf. \cite{Hirota1,Hirota2,Homotopy,QNCL}, this provides us with some new examples of integrable scalar lattice equations. 

Conventionally the Hamiltonian has been the central object in (continuous) integrable systems \cite{Hamiltonian}. Of course, it is often possible to pass between Lagrangian and Hamiltonian theories via Legendre transforms, although this is in many (non-Newtonian) cases not a trivial matter. Nevertheless, most integrable partial differential equations seem to admit a Lagrangian description; in fact, a universal Lagrangian structure for integrable systems admitting a Lax pair was formulated by Zakharov and Mikhailov \cite{Sasha}. In the discrete case one can argue that the Lagrangian is the more fundamental object\footnote{Drawing the parallel with quantum mechanics, we adhere to Dirac's opinion \cite{Dirac}.}, and in fact Lagrangian structures have been established for several discrete integrable systems such as Lagrangian mappings \cite{Veselov88,Veselov,VesMoser}. These Lagrangian descriptions are based on a discrete calculus of variations as developed earlier by Cadzow\cite{Cadzow}, Logan\cite{Logan} and Maeda\cite{Maeda} outside the scope of integrable systems. Furthermore, Lagrangians and/or actions were also constructed for integrable two-dimensional lattice equations, cf. \cite{KdV,GD,ABS}. Discrete Lagrangian systems on arbitrary graphs were proposed in \cite{NovShv}, and a discrete variational complex was set up in \cite{Hydon}.

The usual point of view is that the Lagrangian is a scalar object (or equivalently a volume form), which through the Euler-Lagrange equations provides us with one single equation (i.e. one per component of the dependent variable). In contrast, we take the point of view that in the case of an integrable system, where due to the multidimensional consistency several equations can be imposed simultaneously on one and the same dependent variable, the Lagrangian should actually be an extended object capable of producing a multitude of consistent equations from a variational principle. Thus we propose in this paper an action in which the key ingredient is a Lagrangian 2-form (in the case of integrable discrete equations in two independent variables) or, more generally, a multiform (in the case of a larger number of independent variables). Although the notion of a Lagrangian multiform is not new, and goes back to  Cartan and Lepage \cite{Cartan, Lepage}, cf. also \cite{Kastrup} for a review, even in those theories the role of the Lagrangian is that of a volume form producing the equations of motion in a conventional way. 

In the original ABS paper \cite{ABS}, action functionals were given for the whole list of lattice equations in their classification. In the present paper we reformulate the Lagrangian structures of these lattice systems, identifying a specific form of the Lagrangians, and we show by explicit computation that for each case considered, a closure-type relation holds when we embed these systems into a higher dimensional lattice. This closure relation effectively indicates that the discrete Lagrangian is a closed 2-form on the multidimensional lattice, and we consider this as a manifestation of multidimensional consistency on the Lagrangian level. We argue that since this relation implies surface independence of the relevant action, the variational principle that describes multidimensionally consistent systems requires variations with respect not only to the field variables, but also to the geometry of the lattice on which the integrable discrete equation is defined. By presenting an explicit continuous example of a multidimensionally consistent system, namely the \emph{generating PDE} associated with the lattice KdV equation \cite{SPDE}, we draw in the penultimate section some analogies with the continuous case.

\section{Multidimensional Consistency of Lattices and 3-point Lagrangians}
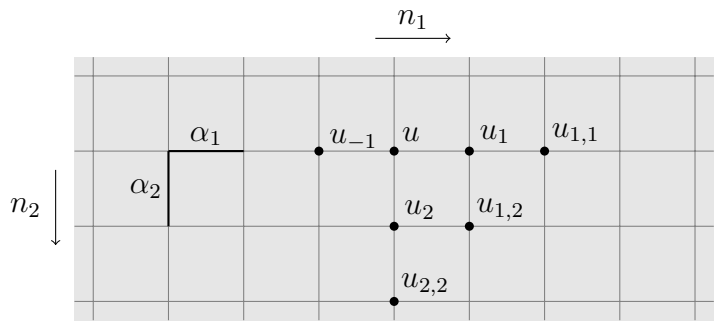
\begin{figure}[h]
  \begin{center}
\begin{tikzpicture}[every node/.style={minimum size=1cm},on grid]
 \fill[gray!20] (0.75,0.75) rectangle (9.25,4.25);
 \draw[black!50, thin] (0.75,0.75) grid (9.25,4.25);
 \draw[thick] (2,3) -- (3,3); \node at (2.5,3.2) {$\al_{1}$};
 \draw[thick] (2,3) -- (2,2); \node at (1.7,2.5) {$\al_{2}$};
 \fill (4,3) circle (0.06) node[above right=-0.5] {$\;\;\;u_{-1}$};
 \fill (5,3) circle (0.06) node[above right=-0.4] {$u$};
 \fill (6,3) circle (0.06) node[above right=-0.4] {$\;\;u_{1}$};
 \fill (7,3) circle (0.06) node[above right=-0.4] {$\;\;u_{1,1}$};
 \fill (5,2) circle (0.06) node[above right=-0.4] {$\;\;u_{2}$};
 \fill (6,2) circle (0.06) node[above right=-0.4] {$\;\;u_{1,2}$};
 \fill (5,1) circle (0.06) node[above right=-0.4] {$\;\;u_{2,2}$};
 \draw[->] (4.75,4.5) -- (5.75,4.5);  \node at (5.25,4.75) {$n_{1}$};
 \draw[->] (0.5,2.75) -- (0.5,1.75); \node at (0.1,2.25) {$n_{2}$};
\end{tikzpicture}
  \end{center}
  \caption{2-d lattice}
  \label{2dlattice}
 \end{figure}

\setcounter{equation}{0}
In this paper, following \cite{ABS}, we consider lattice systems of the following form: there are two independent (discrete) variables $n_{1},n_{2}$ corresponding to two lattice directions, two lattice parameters $\al_{1}$, $\al_{2}$ (which can be thought of as measures for the grid size) associated with the $n_{1}$, $n_{2}$ directions respectively, and a scalar dependent variable $u(n_{1},n_{2})$. The parameters $\al_{1}$, $\al_{2}$ may take on continuous values, whilst $n_{1},n_{2}$ are discrete coordinates of the lattice, but may in principle take on continuous values as long as elementary shifts in these variables amount to increments by one unit.
For ease of notation, let $u=u(n_{1},n_{2})$, $u_{1}=u(n_{1}+1,n_{2})$ and $u_{2}=u(n_{1},n_{2}+1)$, as indicated in Figure \ref{2dlattice}. Backwards shifts in $u$ are denoted by $u_{-1}=u(n_{1}-1,n_{2})$ and $u_{-2}=u(n_{1},n_{2}-1)$.
In \cite{ABS} the classification problem of quadrilateral lattice equations of the following form was considered
 \be\label{Q=0}
  Q(u,u_{1},u_{2},u_{1,2};\al_{1},\al_{2})=0
 \ee
where $Q$ is affine linear and subject to the symmetry of the square (D$_{4}$ symmetry). The main classification criterion is that of multidimensional consistency, by which we mean the following.

\begin{figure}[h]
  \begin{center}
\begin{tikzpicture}[every node/.style={minimum size=1cm},on grid]
 \fill[gray!20] (0,0) rectangle (3,3);
\begin{scope}[every node/.append style={yslant=0.5,xslant=-1},xslant=2]
 \fill[gray!30] (-6,3) rectangle (-3,3.75);
\end{scope}
\begin{scope}[every node/.append style={yslant=0.5},yslant=0.5]
 \fill[gray!40] (3,-1.5) rectangle (4.5,1.5);
\end{scope}
 \draw[gray, very thin] (1.5,0.75) -- (4.5,0.75);
 \draw[gray, very thin] (1.5,0.75) -- (1.5,3.75);
\begin{scope}[every node/.append style={yslant=0.5},yslant=0.5]
 \draw[gray, very thin] (0,0) -- (1.5,0);
\end{scope}
 \draw (0,0) rectangle (3,3);
\begin{scope}[every node/.append style={yslant=0.5,xslant=-1},xslant=2]
 \draw (-6,3) rectangle (-3,3.75);
\end{scope}
\begin{scope}[every node/.append style={yslant=0.5},yslant=0.5]
 \draw (3,-1.5) rectangle (4.5,1.5);
\end{scope}
 \draw (4.5,3.75) circle (0.1) node[above right=-0.3] {$u_{1}$};
 \draw[gray, very thin] (1.5,0.75) circle (0.1) node[above right=-0.3] {$u_{2}$};
 \draw (0,3) circle (0.1) node[left=-0.1] {$u_{3}$};
 \draw (3,0) circle (0.1) node[below right=-0.2] {$u_{1,2,3}$};
 \fill (1.5,3.75) circle (0.1) node[above left=-0.4] {$u$};
 \fill (4.5,0.75) circle (0.1) node[right] {$u_{1,2}$};
 \fill (0,0) circle (0.1) node[below left=-0.3] {$u_{2,3}$};
 \fill (3,3) circle (0.1) node[below right=-0.3] {$u_{1,3}$};
\end{tikzpicture}
  \end{center}
  \caption{Elementary cube}
  \label{cube}
 \end{figure}
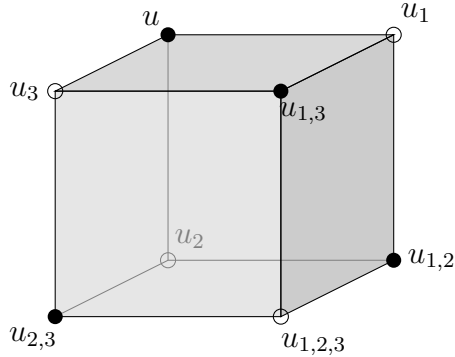

It is known \cite{CAC,B+S} that these equations can be consistently embedded in a 3-dimensional lattice, imposing the same form of the equation (with appropriate parameters) in all 2-dimensional sublattices. This has appeared in many places in the recent literature, but for the sake of self-containedness we present it again here. Imposing in addition to the equation $Q(u,u_{1},u_{2},u_{1,2};\al_{1},\al_{2})=0$ the equations $Q(u,u_{2},u_{3},u_{2,3};\al_{2},\al_{3})=0$ and $Q(u,u_{3},u_{1},u_{1,3};\al_{3},\al_{1})=0$ on the other faces of the elementary cube of Figure \ref{cube}, and given the initial values $u$, $u_{1}$, $u_{2}$, $u_{3}$, there are in principle three ways in which to compute the value of $u_{1,2,3}$. If these three values coincide, the equation is called consistent-around-the-cube, and is considered to be multi-dimensionally consistent. As was pointed out in \cite{NRGO} this phenomenon is analogous to the existence of integrable hierarchies in nonlinear evolution equations of soliton type. 

The result of the classification study of \cite{ABS} was a list of 9 equations, up to M\"obius transformations, labelled H1-H3, Q1-Q4 and A1-A2. Some of these were already well-known, e.g. the discrete KdV equation \cite{Hirota1}, but the classification also produced some new equations. Furthermore, it was shown in \cite{ABS} that all these equations admit an action principle, which is based on the so-called 3-leg form of the quadrilateral equation. In some cases, namely lattice equations ``of KdV type'' a Lagrangian description had been previously established, \cite{KdV,GD}. From the actions given in \cite{ABS}, one can infer 4-point Lagrangians, however, for our purpose it is more useful to identify 3-point Lagrangians. In terms of these the action will take the form 
 \be
  S = \sum_{n_{1},n_{2}\in\mathbb{Z}}{\cL(u,u_{1},u_{2};\al_{1},\al_{2})},
 \ee
and in this specific form Lagrangians of all ABS equations can be established.

The discrete Euler-Lagrange equations arising from the variational principle that $\dl S=0$, under local variations $\dl u(n_{1},n_{2})$ of the dependent variable, are given by
 \be\label{EL}
  \frac{\ptl}{\ptl u}\biggl(\cL(u,u_{1},u_{2};\al_{1},\al_{2})+\cL(u_{-1},u,u_{-1,2};\al_{1},\al_{2})
  +\cL(u_{-2},u_{1,-2},u;\al_{1},\al_{2})\biggr) = 0
 \ee
Below we list specific examples of ABS lattice equations together with their 3-point Lagrangians. Although similar formulae can be established for the remaining cases in the ABS list, we will restrict ourselves here to these particular examples for the remainder of the paper. It should be noted that the discrete Euler-Lagrange equations \eqref{EL} do not give the quadrilateral lattice equations themselves, but rather a discrete derivative of the original equation which is defined on 7 points of the lattice (lattice equations on 7-point stencils have attracted a considerable amount of interest in recent years, cf. e.g. \cite{Maciej}). The Euler-Lagrange equation actually results in a compound of two copies of the 3-leg form of the original equation: one reflected in the $n_{1}$-direction and one reflected in the $n_{2}$-direction\footnote{We note that in the generic case, the D$_{4}$ symmetry is not always manifest on the level of the 3-leg form, e.g. in the case of Q4 where the 3-leg form lives on the level of the uniformizing variables of the relevant elliptic curve. It is through the connection with the affine linear form of the equations that the symmetry under reversal of the shifts becomes apparent.}.

\subsection{H1}
This is the discrete potential Korteweg de Vries equation, one of the most fundamental examples in discrete integrable systems. The Lagrangian was first given in \cite{KdV}.\\
\bse
The original equation is
 \be\label{H1}
  (u-u_{1,2})(u_{1}-u_{2})-\al_{1}+\al_{2}=0;
 \ee
written in 3-leg form this is 
 \be\label{H1leg}
  (u+u_{1}) - (u+u_{2}) = \frac{\al_{1}-\al_{2}}{u-u_{1,2}}
 \ee
and it possesses the Lagrangian
 \be
  \cL = (u_{1}-u_{2})u-(\al_{1}-\al_{2})\ln(u_{1}-u_{2})
 \ee
which through the Euler-Lagrange equation \eqref{EL} leads to the 7-point equation
 \be
   u_{1}-u_{-2}-\frac{\al_{1}-\al_{2}}{u-u_{1,-2}}
 + u_{-1}-u_{2}-\frac{\al_{1}-\al_{2}}{u-u_{-1,2}}= 0
 \ee 
\ese
which contains two copies of the 3-leg form \eqref{H1leg}. We get a similar 7-point equation, consisting of two copies of the original 4-point equation, from all of the Lagrangians given below.

\subsection{H2}
\bse
The original equation is
 \be
  (u-u_{1,2})(u_{1}-u_{2})-(\al_{1}-\al_{2})(u+u_{1}+u_{2}+u_{1,2})-\al_{1}^2+\al_{2}^2=0;
 \ee
written in 3-leg form this is 
 \be
  \frac{u+u_{1}+\al_{1}}{u+u_{2}+\al_{2}}
   = \frac{u-u_{1,2}+\al_{1}-\al_{2}}{u-u_{1,2}-\al_{1}+\al_{2}}
 \ee
and it possesses the Lagrangian
 \bea
  \cL & = & (u+u_{1}+\al_{1})\ln(u+u_{1}+\al_{1})-(u+u_{2}+\al_{2})\ln(u+u_{2}+\al_{2})\nn\\
         && -(u_{1}-u_{2}+\al_{1}-\al_{2})\ln(u_{1}-u_{2}+\al_{1}-\al_{2})\nn\\
         && +(u_{1}-u_{2}-\al_{1}+\al_{2})\ln(u_{1}-u_{2}-\al_{1}+\al_{2}).
 \eea
\ese
 
\subsection{H3}
\bse
This is also known as the discrete modified (potential) KdV equation.\\
The original equation is
 \be
  \al_{1}(uu_{1}+u_{2}u_{1,2})-\al_{2}(uu_{2}+u_{1}u_{1,2})+\dl(\al_{1}^2-\al_{2}^2)=0;
 \ee
written in 3-leg form this is 
 \bea
  \frac{e^{2x+2x_{1}}+\dl e^{2a_{1}}}{e^{2x+2x_{2}}+\dl e^{2a_{2}}}
   = \frac{\sinh(x-x_{1,2}-a_{1}+a_{2})}{\sinh(x-x_{1,2}+a_{1}-a_{2})}
 \eea
where $u=e^{2x}$ and $\al_{1}=e^{2a_{1}}$, and it possesses the Lagrangian \bea
  \cL & = & -\Li_2(\frac{uu_{1}}{-\al_{1}\dl})+\Li_2(\frac{uu_{2}}{-\al_{2}\dl})
            +\Li_2(\frac{\al_{2}u_{1}}{\al_{1}u_{2}})-\Li_2(\frac{\al_{1}u_{1}}{\al_{2}u_{2}})\nn\\
         && +\ln\biggl(\frac{\al_{1}^2}{\al_{2}^2}\biggr)\ln(u)+\ln(\al_{2}^2)\ln\biggl(\frac{u_{1}}{u_{2}}\biggr)
 \eea
where $\Li_2(z)$ is the dilogarithm function given in \eqref{dilog} of appendix A, where some useful identities for the dilogarithm function are presented. 
\ese

\subsection{Q1$\mid_{\dl=0}$}
\bse
The original equation is 
 \be
  \al_{1}(u-u_{2})(u_{1}-u_{1,2})-\al_{2}(u-u_{1})(u_{2}-u_{1,2})=0;
 \ee
written in 3-leg form this is 
 \be
  \frac{\al_{1}}{u-u_{1}}-\frac{\al_{2}}{u-u_{2}} = \frac{\al_{1}-\al_{2}}{u-u_{1,2}}
 \ee and it possesses the Lagrangian
 \be
  \cL = \al_{1}\ln(u-u_{1})-\al_{2}\ln(u-u_{2})-(\al_{1}-\al_{2})\ln(u_{1}-u_{2}).
 \ee
\ese
 
\subsection{Q1$\mid_{\dl\neq 0}$}
\bse
The original equation is
 \be
  \al_{1}(u-u_{2})(u_{1}-u_{1,2})-\al_{2}(u-u_{1})(u_{2}-u_{1,2})+\dl^2\al_{1}\al_{2}(\al_{1}-\al_{2})=0;
 \ee
written in 3-leg form this is 
 \be
  \biggl(\frac{u-u_{1}+\al_{1}\dl}{u-u_{1}-\al_{1}\dl}\biggr)
  \biggl(\frac{u-u_{2}-\al_{2}\dl}{u-u_{2}+\al_{2}\dl}\biggr)
   = \biggl(\frac{u-u_{1,2}+\al_{1}\dl-\al_{2}\dl}{u-u_{1,2}-\al_{1}\dl+\al_{2}\dl}\biggr)
 \ee
and it possesses the Lagrangian
 \bea
\fl \cL & = & (u-u_{1}+\al_{1}\dl)\ln(u-u_{1}+\al_{1}\dl)-(u-u_{1}-\al_{1}\dl)\ln(u-u_{1}-\al_{1}\dl)\nn\\
\fl        && -(u-u_{2}+\al_{2}\dl)\ln(u-u_{2}+\al_{2}\dl)+(u-u_{2}-\al_{2}\dl)\ln(u-u_{2}-\al_{2}\dl)\nn\\
\fl        && -(u_{1}-u_{2}+\al_{1}\dl-\al_{2}\dl)\ln(u_{1}-u_{2}+\al_{1}\dl-\al_{2}\dl)\nn\\
\fl        && +(u_{1}-u_{2}-\al_{1}\dl+\al_{2}\dl)\ln(u_{1}-u_{2}-\al_{1}\dl+\al_{2}\dl).
 \eea         
\ese
  
\subsection{Q3$\mid_{\dl=0}$}
\bse
Written in a slightly different form, this equation is known as the Homotopy equation, and appears in the literature in \cite{Homotopy}.
The original equation is
 \be
  (\al_{2}^2-\al_{1}^2)(uu_{1,2}+u_{1}u_{2})+\al_{2}(\al_{1}^2-1)(uu_{1}+u_{2}u_{1,2})
  -\al_{1}(\al_{2}^2-1)(uu_{2}+u_{1}u_{1,2})=0;
 \ee
written in 3-leg form this is 
 \bea
  \biggl(\frac{\sinh(x-x_{1}+a_{1})}{\sinh(x-x_{1}-a_{1})}\biggr)
  \biggl(\frac{\sinh(x-x_{2}-a_{2})}{\sinh(x-x_{2}+a_{2})}\biggr)
   = \biggl(\frac{\sinh(x-x_{1,2}+a_{1}-a_{2})}{\sinh(x-x_{1,2}-a_{1}+a_{2})}\biggr)&&\nn\\
   &&
 \eea
where $u=e^{2x}$ and $\al_{1}=e^{2a_{1}}$, and it possesses the Lagrangian
 \bea
  \cL & = & -\Li_2\biggl(\frac{\al_{1}  u}{u_{1}}\biggr)+\Li_2\biggl(\frac{u}{\al_{1}u_{1}}\biggr)
                   +\Li_2\biggl(\frac{\al_{2} u}{u_{2}}\biggr)-\Li_2\biggl(\frac{u}{\al_{2}u_{2}}\biggr)\nn\\
                && +\Li_2\biggl(\frac{\al_{1}u_{1}}{\al_{2}u_{2}}\biggr)-\Li_2\biggl(\frac{\al_{2}u_{1}}{\al_{1}u_{2}}\biggr)
                   +\ln(\al_{1}^2)\ln\biggl(\frac{\al_{2}u_{1}}{\al_{1}u_{2}}\biggr).
 \eea
\ese

\subsection{A1}
\bse
The original equation is
 \be
  \al_{1}(u+u_{2})(u_{1}+u_{1,2})-\al_{2}(u+u_{1})(u_{2}+u_{1,2})-\dl^2\al_{1}\al_{2}(\al_{1}-\al_{2})=0;
 \ee
written in 3-leg form this is 
 \be
  \biggl(\frac{u+u_{1}+\al_{1}\dl}{u+u_{1}-\al_{1}\dl}\biggr)
  \biggl(\frac{u+u_{2}-\al_{2}\dl}{u+u_{2}+\al_{2}\dl}\biggr)
   = \biggl(\frac{u-u_{1,2}+\al_{1}\dl-\al_{2}\dl}{u-u_{1,2}-\al_{1}\dl+\al_{2}\dl}\biggr)
 \ee
and it possesses the Lagrangian
 \bea
  \cL & = & (u+u_{1}+\al_{1}\dl)\ln(u+u_{1}+\al_{1}\dl)-(u+u_{1}-\al_{1}\dl)\ln(u+u_{1}-\al_{1}\dl)\nn\\
         && -(u+u_{2}+\al_{2}\dl)\ln(u+u_{2}+\al_{2}\dl)+(u+u_{2}-\al_{2}\dl)\ln(u+u_{2}-\al_{2}\dl)\nn\\
         && -(u_{2}-u_{1}+\al_{1}\dl-\al_{2}\dl)\ln(u_{2}-u_{1}+\al_{1}\dl-\al_{2}\dl)\nn\\
         && +(u_{2}-u_{1}-\al_{1}\dl+\al_{2}\dl)\ln(u_{2}-u_{1}-\al_{1}\dl+\al_{2}\dl).
 \eea       
\ese
  
\subsection{A2}
\bse
The original equation is
 \be
  (\al_{2}^2-\al_{1}^2)(uu_{1}u_{2}u_{1,2}+1)+\al_{2}(\al_{1}^2-1)(uu_{2}+u_{1}u_{1,2})
  -\al_{1}(\al_{2}^2-1)(uu_{1}+u_{2}u_{1,2})=0;
 \ee
written in 3-leg form this is 
 \bea
  \biggl(\frac{\sinh(x+x_{1}+a_{1})}{\sinh(x+x_{1}-a_{1})}\biggr)
  \biggl(\frac{\sinh(x+x_{2}-a_{2})}{\sinh(x+x_{2}+a_{2})}\biggr)
   = \biggl(\frac{\sinh(x-x_{1,2}+a_{1}-a_{2})}{\sinh(x-x_{1,2}-a_{1}+a_{2})}\biggr)&&\nn\\
   &&
 \eea
where $u=e^{2x}$ and $\al_{1}=e^{2a_{1}}$, and it possesses the Lagrangian
 \bea
  \cL & = & -\Li_2(\al_{1} uu_{1})+\Li_2\biggl(\frac{uu_{1}}{\al_{1}}\biggr)
            +\Li_2(\al_{2} uu_{2})-\Li_2\biggl(\frac{uu_{2}}{\al_{2}}\biggr)\nn\\
         && +\Li_2\biggl(\frac{\al_{1}u_{2}}{\al_{2}u_{1}}\biggr)-\Li_2\biggl(\frac{\al_{2}u_{2}}{\al_{1}u_{1}}\biggr)
            +\ln(\al_{1}^2)\ln\biggl(\frac{\al_{2}u_{2}}{\al_{1}u_{1}}\biggr).
 \eea
\ese 

For the Lagrangians of the cases given in the list 2.1-2.8, we will next establish an important new property.
 
\section{Closure relation and Lagrangian 2-forms}
\setcounter{equation}{0} 
The main observation of this paper is that all the lattice systems, together with their 3-point Lagrangians, as given in the previous section, possess a remarkable property which we refer to as the \emph{closure relation}, when we embed both the equation and the Lagrangian in a 3-dimensional lattice. In order to formulate this property we introduce the notation of the difference operator $\Dl_{i}$ which acts on functions $f$ of $u=u(n_{1},n_{2},n_{3})$ by the formula $\Dl_{i}f(u)=f(u_{i})-f(u)$, and on a function $g$ of $u$ and its shifts by the formula $\Dl_{i}g(u,u_{j},u_{k})=g(u_{i},u_{i,j},u_{i,k})-g(u,u_{j},u_{k})$, in which, as before, the suffix $i$ denotes a shift in the direction associated with the variable $n_{i}$. The following statement holds true.\\[10pt]
\textbf{Proposition:}\\
\emph{All the 3-point Lagrangians given in the list 2.1-2.8 when embedded in a three-dimensional lattice, as explained in section 2, satisfy the following relation on solutions of the quadrilateral lattice system:}
 \be\label{dclos}
  \Dl_{1}\cL(u,u_{2},u_{3};\al_{2},\al_{3})+\Dl_{2}\cL(u,u_{3},u_{1};\al_{3},\al_{1})
  +\Dl_{3}\cL(u,u_{1},u_{2};\al_{1},\al_{2}) = 0
 \ee
\paragraph{}
This can be established by explicit computation, and has been verified in all cases in the list 2.1-2.8. Below we will demonstrate this computation for the case of H1. Furthermore, in appendix B we will present the computation in the case of H3, which is somewhat more involved and relies on a number of identities for the dilogarithm function $\Li_{2}$, see e.g. \cite{Lewin,Kirillov}, the relevant ones of which have been reproduced in appendix A. For the Lagrangians of the remaining equations in the ABS list the computations are more implicit and we delegate those to a future publication.

\paragraph{Example: H1}
To illustrate the proposition in the simplest case, we perform the following computation. By definition of the Lagrangians we have
 \bea\label{H1closa}
\fl    &&  \Dl_{1}\cL(u,u_{2},u_{3};\al_{2},\al_{3})+\Dl_{2}\cL(u,u_{3},u_{1};\al_{3},\al_{1})
        +\Dl_{3}\cL(u,u_{1},u_{2};\al_{1},\al_{2})\nn\\
\fl & = &  (u_{1,2}-u_{1,3})u_{1}-(\al_{2}-\al_{3})\ln(u_{1,2}-u_{1,3})-(u_{2}-u_{3})u\nn\\
\fl    && +(\al_{2}-\al_{3})\ln(u_{2}-u_{3})+(u_{2,3}-u_{1,2})u_{2} -(\al_{3}-\al_{1})\ln(u_{2,3}-u_{1,2})\nn\\
\fl    && -(u_{3}-u_{1})u+(\al_{3}-\al_{1})\ln(u_{3}-u_{1})+(u_{1,3}-u_{2,3})u_{3}\nn\\
\fl    && -(\al_{1}-\al_{2})\ln(u_{1,3}-u_{2,3})-(u_{1}-u_{2})u+(\al_{1}-\al_{2})\ln(u_{1}-u_{2}).
 \eea
Noting that the differences between the double-shifted terms have the form 
 \bea
  u_{1,2}-u_{1,3} & = & -\frac{(\al_{2}-\al_{3})u_{1}+(\al_{3}-\al_{1})u_{2}+(\al_{1}-\al_{2})u_{3}}
                             {(u_{1}-u_{2})(u_{2}-u_{3})(u_{3}-u_{1})}(u_{2}-u_{3})\nn\\
                   & = & A_{1,2,3}(u_{2}-u_{3})
 \eea
where $A_{1,2,3}$ is invariant under permutations of the indices, the expression \eqref{H1closa} reduces to
 \bea\label{H1clos}
\fl A_{1,2,3}(u_{2}-u_{3})u_{1}-(\al_{2}-\al_{3})\ln\bigl(A_{1,2,3}(u_{2}-u_{3})\bigr)
    -(u_{2}-u_{3})u+(\al_{2}-\al_{3})\ln(u_{2}-u_{3})&&\nn\\
\fl +A_{1,2,3}(u_{3}-u_{1})u_{2}-(\al_{3}-\al_{1})\ln\bigl(A_{1,2,3}(u_{3}-u_{1})\bigr)
    -(u_{3}-u_{1})u+(\al_{3}-\al_{1})\ln(u_{3}-u_{1})&&\nn\\
\fl +A_{1,2,3}(u_{1}-u_{2})u_{3}-(\al_{1}-\al_{2})\ln\bigl(A_{1,2,3}(u_{1}-u_{2})\bigr)
    -(u_{1}-u_{2})u+(\al_{1}-\al_{2})\ln(u_{1}-u_{2})&&\nn\\
\fl = 0&&
\eea  
where we have tried to organize the succession of terms to make it manifest which groupings of terms cancel out against each other.

In a similar way the cases of H2, Q1 and A1 can be verified, whereas the cases of H3, Q3$|_{\dl=0}$ and A2 all involve the dilogarithm function $\Li_{2}$ and can be verified along similar lines as the computation in appendix B.

\begin{figure}[h]
\begin{center}
\begin{tikzpicture}[every node/.style={minimum size=1cm}]
\begin{scope}[every node/.append style={yslant=-0.5},yslant=-0.5]
  \filldraw[fill=gray!80] (0.5,0.5) rectangle (1,1);
  \filldraw[fill=gray!80] (1.5,1.5) rectangle (2,2);
\end{scope}
\begin{scope}[every node/.append style={yslant=0.5,xslant=-1},yslant=0.5,xslant=-1]
  \filldraw[fill=gray!20] (0,-0.5) rectangle (0.5,0);
  \filldraw[fill=gray!20] (1,-0.5) rectangle (1.5,0);
  \filldraw[fill=gray!20] (2,0.5) rectangle (2.5,1);
  \filldraw[fill=gray!20] (2.5,-1.5) rectangle (2,-1);
\end{scope}
\begin{scope}[every node/.append style={yslant=-0.5},yslant=-0.5]
  \filldraw[fill=gray!80] (0.5,0) rectangle (1,0.5); \draw[gray,very thin] (0.5,0) -- (1,0.5);
  \filldraw[fill=gray!80] (1.5,0.5) rectangle (2,1); \draw[gray,very thin] (1.5,0.5) -- (2,1);
  \filldraw[fill=gray!80] (2,0.5) rectangle (2.5,1); \draw[gray,very thin] (2,0.5) -- (2.5,1);
  \filldraw[fill=gray!80] (1.5,1) rectangle (2,1.5); \draw[gray,very thin] (1.5,1) -- (2,1.5);
  \filldraw[fill=gray!80] (2.5,1.5) rectangle (3,2); \draw[gray,very thin] (2.5,1.5) -- (3,2);
  \filldraw[fill=gray!80] (2,2) rectangle (2.5,2.5); \draw[gray,very thin] (2,2) -- (2.5,2.5);
  \filldraw[fill=gray!80] (1.5,2) rectangle (2,2.5); \draw[gray,very thin] (1.5,2) -- (2,2.5);
  \filldraw[fill=gray!80] (1,1.5) rectangle (1.5,2); \draw[gray,very thin] (1,1.5) -- (1.5,2);
\end{scope}
\begin{scope}[every node/.append style={yslant=0.5},yslant=0.5]
  \filldraw[fill=gray!50] (1,-1) rectangle (1.5,-0.5); \draw[gray,very thin] (1,-0.5) -- (1.5,-1);
  \filldraw[fill=gray!50] (2.5,-2) rectangle (3,-1.5); \draw[gray,very thin] (2.5,-1.5) -- (3,-2);
  \filldraw[fill=gray!50] (3,-2) rectangle (3.5,-1.5); \draw[gray,very thin] (3,-1.5) -- (3.5,-2);
  \filldraw[fill=gray!50] (3,-1.5) rectangle (3.5,-1); \draw[gray,very thin] (3,-1) -- (3.5,-1.5);
  \filldraw[fill=gray!50] (2,-1) rectangle (2.5,-0.5); \draw[gray,very thin] (2,-0.5) -- (2.5,-1);
  \filldraw[fill=gray!50] (2.5,-0.5) rectangle (3,0); \draw[gray,very thin] (2.5,0) -- (3,-0.5);
\end{scope}
\begin{scope}[every node/.append style={yslant=0.5,xslant=-1},yslant=0.5,xslant=-1]
  \filldraw[fill=gray!20] (0.5,-0.5) rectangle (1,0); \draw[gray,very thin] (0.5,-0.5) -- (1,0);
  \filldraw[fill=gray!20] (1,0) rectangle (1.5,0.5); \draw[gray,very thin] (1,0) -- (1.5,0.5);
  \filldraw[fill=gray!20] (2.5,0.5) rectangle (3,1); \draw[gray,very thin] (2.5,0.5) -- (3,1);
  \filldraw[fill=gray!20] (2.5,0) rectangle (3,0.5); \draw[gray,very thin] (2.5,0) -- (3,0.5);
  \filldraw[fill=gray!20] (1.5,-0.5) rectangle (2,0); \draw[gray,very thin] (1.5,-0.5) -- (2,0);
  \filldraw[fill=gray!20] (2,-1) rectangle (2.5,-0.5); \draw[gray,very thin] (2,-1) -- (2.5,-0.5);
  \filldraw[fill=gray!20] (1,-1.5) rectangle (1.5,-1); \draw[gray,very thin] (1,-1.5) -- (1.5,-1);
  \filldraw[fill=gray!20] (0,-1.5) rectangle (0.5,-1);
\end{scope}
\begin{scope}[every node/.append style={yslant=0.5,xslant=-1},yslant=0.5,xslant=-1]
 \draw[very thick] (0.5,0) -- (0,0);
 \draw[very thick] (0,0) -- (0,-1.5);
 \draw[very thick] (0,-1.5) -- (0.5,-1.5);
 \draw[very thick] (0.5,-1.5) -- (0.5,-2);
 \draw[very thick] (0.5,-2) -- (1.5,-2);
 \draw[very thick] (1.5,-2) -- (2,-1.5);
 \draw[very thick] (2,-1.5) -- (2.5,-1.5);
 \draw[very thick] (2.5,-1.5) -- (2.5,-1);
\end{scope}
\end{tikzpicture}
\caption{Example of a surface with boundary}
\label{surface}
\end{center}
\end{figure}
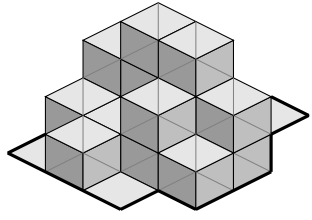

\paragraph{}
In order to discuss the implications of the statement above, we need to introduce some further notation. Let $\bde_{i}$ denote the unit vector in the lattice direction labelled by $i$ and let any point in the multidimensional lattice be specified by the vector $\bn$ whose components are the coordinates $n_{1},n_{2},\dots$ of the lattice\footnote{If we select a lattice of finite dimensionality we could write the coordinates on the lattice as $\bn=(n_{1},n_{2},\dots)$, where the lattice directions are labelled according to the natural numbers. However in principle one could also have an infinite dimensional lattice and even a lattice labelled by an uncountable set.}, then elementary shifts in the lattice can be generated by the action $\bn\rightarrow\bn+\bde_{i}$. Specifying an elementary oriented plaquette in this lattice requires the following data: the position $\bn$ of one of its vertices in the lattice and the lattice directions given by the base vectors $\bde_{i},\bde_{j}$. One way to characterize the oriented plaquette is by the ordered triplet $\sigma_{ij}(\bn)=(\bn,\bn+\bde_{i},\bn+\bde_{j})$. Since the 3-point Lagrangians depend on two directions in the lattice, and when embedded in a multidimensional lattice at each point can be associated with an oriented plaquette $\sigma_{ij}(\bn)$, we can think of these Lagrangians as defining a discrete 2-form $\cL_{ij}(\bn)$ whose evaluation on that plaquette is given by the Lagrangian function as follows
 \be\label{cL}
  \cL_{ij}(\bn) = \cL(u(\bn),u(\bn+\bde_{i}),u(\bn+\bde_{j});\al_{i},\al_{j}).
 \ee
Here it is understood that \eqref{cL} is the contribution to the action functional associated with the given plaquette $\sigma_{ij}(\bn)$ as described above. Choosing now a surface $\sigma$ in the multidimensional lattice consisting of a connected configuration of elementary plaquettes $\sigma_{ij}(\bn)$, such as illustrated in Figure \ref{surface} (which could be an infinite surface or a compact surface, with or without boundary)  
we can define an action on that surface simply by summing up the contributions $\cL_{ij}$ from each of the plaquettes on the surface, taking into account the directions associated with each face, i.e. we perform the sum:
 \be\label{S}
  S = S[u(\bn);\sigma] = \sum_\sigma{\cL} = \sum_{\sigma_{ij}(\bn)\in\sigma}{\cL_{ij}(\bn)}.
 \ee
The sum in \eqref{S} is unambiguous for two reasons: first, because all the Lagrangians considered in 2.1-2.8 have the property of antisymmetry up to a constant with respect to transformations $i\leftrightarrow j$, i.e. $\cL_{ij}(\bn)=-\cL_{ji}(\bn)+\textrm{constant}$; second, we choose the base point $\bn$ in such a way that $\cL_{ij}(\bn)$, defined on $\sg_{ij}(\bn)$, involves $u(\bn)$ along with its shifts only in the positive $i$ and $j$ directions. We choose throughout this paper not to use the abstract notation of difference forms, cf. e.g. \cite{Hydon2}, because we want to demonstrate on the basis of the examples given that all statements can be established through concrete computations.

\begin{figure}[h]
\begin{center}
\begin{tikzpicture}[every node/.style={minimum size=1cm},on grid]
\begin{scope}[every node/.append style={yslant=0.5,xslant=-1},yslant=0.5,xslant=-1]
 \fill[gray!20] (0,0) rectangle (2,2);
 \draw[step=0.5] (0,0) grid (2,2);
\end{scope}
\end{tikzpicture}
\begin{tikzpicture}[every node/.style={minimum size=1cm},on grid]
\begin{scope}[every node/.append style={yslant=0.5,xslant=-1},yslant=0.5,xslant=-1]
 \fill[gray!20] (0,0) rectangle (2,2);
 \draw[step=0.5] (0,0) grid (2,2);
\end{scope}
\begin{scope}[every node/.append style={yslant=-0.5},yslant=-0.5]
 \filldraw[fill=gray!80] (-1,0.5) rectangle (-0.5,1);
\end{scope}
\begin{scope}[every node/.append style={yslant=0.5},yslant=0.5]
 \filldraw[fill=gray!50] (-0.5,1) rectangle (0,1.5);
\end{scope}
\begin{scope}[every node/.append style={yslant=0.5,xslant=-1},yslant=0.5,xslant=-1]
 \filldraw[fill=gray!20] (1,1.5) rectangle (1.5,2);
 \draw[step=0.5,gray,very thin] (0,0) grid (2,2);
\end{scope}
\end{tikzpicture}
\caption{Local deformation of a discrete surface $\sigma$ to a surface $\sigma'$}
\label{latdef}
\end{center}
\end{figure}
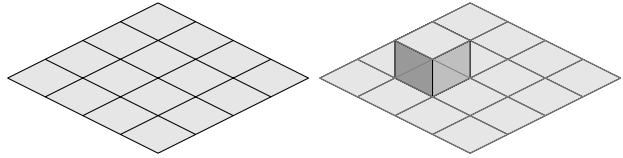

It is obvious from \eqref{S} that the geometry of the surface $\sigma$ forms an integral part of the action functional. The closure relation \eqref{dclos} implies the invariance of the action under local deformations of the surface $\sigma$ while fixing its boundary. This we can easily see by considering an elementary variation of a locally flat surface at a single plaquette illustrated by Figure \ref{latdef}.
If $S$ is the value of the action functional for the undeformed surface in Figure \ref{latdef} the value for the deformed surface in Figure \ref{latdef} can be computed as follows
 \bea\label{Sprime}
\fl  S' & = &  S-\cL(u,u_{i},u_{j};\al_{i},\al_{j})+\cL(u_{k},u_{i,k},u_{j,k};\al_{i},\al_{j})
              +\cL(u_{i},u_{i,j},u_{i,k};\al_{j},\al_{k})\nn\\
\fl        && +\cL(u_{j},u_{j,k},u_{i,j};\al_{k},\al_{i})-\cL(u,u_{j},u_{k};\al_{j},\al_{k})
              -\cL(u,u_{k},u_{i};\al_{k},\al_{i})
 \eea
taking into account the orientation of the deformation $\sigma\rightarrow\sigma'$, defined as a transition between two collections of oriented plaquettes as indicated by the figure. From this argument it follows that the independence of the action under such a deformation is locally equivalent to the closure relation \eqref{dclos}. We consider this invariance an essential aspect of the relevant variational principle underlying multidimensionally consistent lattice systems.

The aim of a Lagrangian multiform description over the usual scalar Lagrangian one is that it should provide us with not just one variational equation, but in principle an arbitrary number of compatible equations. At this stage it is not entirely clear what is the optimal formulation for such a principle, in view of the fact that the closure relations we have established for specific examples rely on the quadrilateral lattice equations themselves, but we will make an attempt in this direction by posing the following.

\paragraph{Discrete variational principle for integrable lattice systems:}
The functions $u(\bn)$ solving an integrable multidimensional lattice system on each discrete quadrilateral surface $\sigma$ are those for which the action $S[u(\bn);\sigma]$ of \eqref{S} is invariant under local deformations of the lattice, as described above, and for which the action attains an extremum under infinitesimal local deformations of the dependent variable $u(\bn)$.

\paragraph{}
The mechanism that we have in mind is as follows. Starting with an action functional $S[u(\bn);\sigma]$ as in \eqref{S} we impose surface independence of this action. This allows us to deform the surface $\sigma$ as we choose, whilst keeping the boundary in place if there is a boundary. Thus, we can always render it into a locally flat surface away from the boundary, where we can choose any pair of local coordinates $n_{i},n_{j}$. In that part of the surface we can then apply the usual variational principle with respect to the field variables $u(\bn)$, leading in the usual manner to the Euler-Lagrange equations in those lattice directions. If these equations subsequently imply the validity of the closure relation for the Lagrangian in terms of which the action is defined, this then ensures that those equations are consistent with invariance of the action under deformation of the surface which in turn allowed the derivation of those equations in the first place. This principle goes farther than just providing a variational derivation of equations of the motion from a given Lagrangian, in that in some sense it also imposes conditions on the class of \emph{admissible Lagrangians} to which this principle applies. What is not clear at this stage is to what extent admissible Lagrangians can be constructed by application of this principle and we do not yet have a general proof that this procedure will automatically lead to multidimensionally consistent lattice equations.

\section{Discussion and comparison with the continuous case}
\setcounter{equation}{0} 
The discrete variational principle formulated in the previous section brings in the geometry as a variable of the action functional. This contrasts starkly with the usual variational principle, where the Euler-Lagrange equations provide information rather on the parametrization of the underlying geometry than on the geometry itself. For instance, in the elementary case of a mechanical system with one degree of freedom the action 
 \be
  S[q(t)] = \int_0^T{L(q,\dot{q},t)dt}
 \ee
contains hardly any geometry at all, but the relevant Lagrange equation tells us how the one-dimensional motion is parametrized in a specific way according to the equations of motion. When we have more than one degree of freedom there is obviously room for nontrivial phase space geometry, but again the variational equations tell us more about how the geometry is parametrized rather than bringing in the geometry as a variational variable. Even in classical string theory \cite{string}, the geometry of the string trajectories (which sweep out a surface in configuration space) plays a role at the level of the dependent variables of the string action rather than of the independent variables which parametrize the surface. In contrast, our proposal involves the geometry of the space of independent variables which is somewhat reminiscent of the de Donder-Weyl formalism \cite{Kastrup}, although in this approach the connection to integrability is not evident. As far as we are aware, all Lagrangian descriptions of (continuous) integrable systems so far involve conventional scalar Lagrangians, even where an attempt is made to give a multi-Lagrangian description of integrable hierarchies, cf. \cite{Nutku,NutPav,BusHoj}.

At the continuous level, rather than systems of partial difference equations we would be looking for systems of partial differential equations. An interesting example of a linear system of PDEs which are mutually compatible is given by the following nonautonomous set of equations.
 \be\label{cLKdV}
  \ptl_{p_{i}}\ptl_{p_{j}}(p_{i}^2-p_{j}^2)\ptl_{p_{i}}\ptl_{p_{j}}w = 4(n_{j}\ptl_{p_{i}}-n_{i}\ptl_{p_{j}})
                                 \frac{1}{p_{i}^2-p_{j}^2}(n_{j}p_{i}^2\ptl_{p_{i}}-n_{i}p_{j}^2\ptl_{p_{j}})w
 \ee
where $i,j$ run over some index set $I$.
Each of these, for fixed labels $i,j$, arise as Euler-Lagrange equations from the Lagrange density
 \be\label{cll}
  \cL_{ij} = \frac{1}{n_{j}n_{i}}\biggl(\frac{1}{2}
                                 (p_{i}^2-p_{j}^2)w_{p_{i}p_{j}}^2+(n_{j}^2w_{p_{i}}^2-n_{i}^2w_{p_{j}}^2)
                                +\frac{p_{i}^2+p_{j}^2}{p_{i}^2-p_{j}^2}(n_{j}w_{p_{i}}-n_{i}w_{p_{j}})^2\biggr)
 \ee
where notably the independent variables $p_{i},p_{j}$ are on an equal footing. Here $w=w(p_{i},p_{j})$ is the dependent scalar variable and the $n_{i},n_{j}$ are a pair of parameters of the equation, where we associate the parameter $n_{i}$ with the variable $p_{i}$, and the parameter $n_{j}$ with the variable $p_{j}$. It conspires that the system of PDEs \eqref{cLKdV}, when the labels $i,j$ are assumed to run over some index set of cardinality larger than 2, is multidimensionally consistent in a similar way as the lattice equations considered in section 2. Furthermore, the Lagrangian \eqref{cll} obeys a closure relation of the following form
 \be\label{cclos}
  \ptl_{p_{i}}\cL_{jk} + \ptl_{p_{j}}\cL_{ki} + \ptl_{p_{k}}\cL_{ij} = 0
 \ee
provided one or the other of the following two relations hold
 \bse
 \be\label{cLKdVa}
  w_{p_{i}p_{j}p_{k}} = -4\biggl(\frac{n_{j}n_{k}p_{i}^2w_{p_{i}}}{(p_{k}^2-p_{i}^2)(p_{i}^2-p_{j}^2)}
                                +\frac{n_{k}n_{i}p_{j}^2w_{p_{j}}}{(p_{i}^2-p_{j}^2)(p_{j}^2-p_{k}^2)}
                                +\frac{n_{i}n_{j}p_{k}^2w_{p_{k}}}{(p_{j}^2-p_{k}^2)(p_{k}^2-p_{i}^2)}\biggr)
 \ee
or
 \be\label{cLKdVb}
   \frac{(p_{i}^2-p_{j}^2)w_{p_{i}p_{j}}}{n_{i}n_{j}}+\frac{(p_{j}^2-p_{k}^2)w_{p_{j}p_{k}}}{n_{j}n_{k}}
  +\frac{(p_{k}^2-p_{i}^2)w_{p_{k}p_{i}}}{n_{k}n_{i}}=0.
 \ee
 \ese
We can infer from the closure relation that the action on the solutions of the system
 \be
  S[w;\sg] = \int_{\sg}\sum_{i,j\in I}{\cL_{ij}dp_{i}\wedge dp_{j}}
 \ee
is independent of the surface $\sg$, relying on Stokes' theorem, and similar to the discrete case described in section 3 by locally rectifying the surface, i.e. deforming it locally to a plane in terms of selected independent variables $p_{i},p_{j}$, we can derive from the Euler-Lagrange equations in those variables the system of PDEs. It seems somewhat artificial in this example to invoke the additional equations \eqref{cLKdVa} and \eqref{cLKdVb}, the need for which is mainly due to the fact that we are dealing with higher order PDEs in terms of the derivatives. We note, however, that the PDEs \eqref{cLKdV}, \eqref{cLKdVa} and \eqref{cLKdVb} all hold true on a large class of solutions given by the Fourier-type integral of the form
 \be\label{w}
  w = \int_{C}{dk\;\;c(k)\prod_{i\in I}\biggl(\frac{p_{i}+k}{p_{i}-k}\biggr)^{n_{i}}}
 \ee
over some suitably chosen curve $C$ in the complex plane and suitably chosen coefficient function $c(k)$, where $I$ denotes the index set as above. This example is inspired by the canonical form of the plane wave factors, i.e. discrete exponential functions, appearing in the solutions of the lattice equations \cite{Homotopy,QNCL}, which explains the use of the notation $p_{i}$ as independent variables for historic reasons. The situation described here is the obvious continuous analogue of the situation described in section 3 of the relation between the closure relation and multidimensional consistency, apart from the fact that we are dealing here with a set of linear equations rather than nonlinear ones.

The full nonlinear case analogous to \eqref{cLKdV} appeared first in \cite{SPDE}, and it represents the full KdV hierarchy as a so-called generating PDE given as follows
 \bea\label{cKdV}
\fl  U_{t_{i}t_{i}t_{j}t_{j}} & = & U_{t_{i}t_{i}t_{j}}\biggl(\frac{1}{t_{i}-t_{j}}+\frac{U_{t_{i}t_{j}}}{U_{t_{i}}}
                                                             +\frac{U_{t_{j}t_{j}}}{U_{t_{j}}}\biggr)
                                   +U_{t_{i}t_{j}t_{j}}\biggl(\frac{1}{t_{j}-t_{i}}+\frac{U_{t_{i}t_{j}}}{U_{t_{j}}}
                                                             +\frac{U_{t_{i}t_{i}}}{U_{t_{i}}}\biggr)\nn\\
\fl  && +U_{t_{i}t_{i}}\biggl(\frac{n_{i}^2}{(t_{i}-t_{j})^2}\frac{U_{t_{j}}^2}{U_{t_{i}}^2}
                             -\frac{U_{t_{i}t_{j}}^2}{U_{t_{i}}^2}-\frac{1}{t_{i}-t_{j}}\frac{U_{t_{i}t_{j}}}{U_{t_{i}}}\biggr)
        -U_{t_{i}t_{j}}\frac{U_{t_{i}t_{i}}U_{t_{j}t_{j}}}{U_{t_{i}}U_{t_{j}}}\nn\\
\fl  && +U_{t_{j}t_{j}}\biggl(\frac{n_{j}^2}{(t_{i}-t_{j})^2}\frac{U_{t_{i}}^2}{U_{t_{j}}^2}
                         -\frac{U_{t_{i}t_{j}}^2}{U_{t_{j}}^2}-\frac{1}{t_{j}-t_{i}}\frac{U_{t_{i}t_{j}}}{U_{t_{j}}}\biggr)\nn\\
\fl  && +\frac{n_{i}^2}{2(t_{i}-t_{j})^3}\frac{U_{t_{j}}}{U_{t_{i}}}(U_{t_{i}}+U_{t_{j}}
                                           +2(t_{j}-t_{i})U_{t_{i}t_{j}})\\
\fl  && +\frac{n_{j}^2}{2(t_{j}-t_{i})^3}\frac{U_{t_{i}}}{U_{t_{j}}}(U_{t_{j}}+U_{t_{i}}
              +2(t_{i}-t_{j})U_{t_{i}t_{j}})
              +\frac{1}{2(t_{i}-t_{j})}U_{t_{i}t_{j}}^2\bigg(\frac{1}{U_{t_{i}}}-\frac{1}{U_{t_{j}}}\biggr),\nn
 \eea
which represents a generalization of the Ernst-Weyl equation of general relativity as was shown in \cite{Xenitidis}. The variables $t_{i}$ are closely related to the $p_{i}$ of the previous linear example, namely by $t_{i}=p_{i}^2$.
It was argued in \cite{SPDE} that \eqref{cKdV} constitutes a multidimensionally consistent system in the same way as the linear equation.
The Lagrangian for equation \eqref{cKdV} is 
 \be
  \cL_{ij} = \frac{1}{2}(t_{i}-t_{j})\frac{U_{t_{i}t_{j}}^2}{U_{t_{i}}U_{t_{j}}}+\frac{1}{2(t_{i}-t_{j})}\biggl(n_{j}^2\frac{U_{t_{i}}}{U_{t_{j}}}+n_{i}^2\frac{U_{t_{j}}}{U_{t_{i}}}\biggr).
 \ee
This satisfies the closure relation \eqref{cclos} provided again that one of two relations hold,
 \bse
 \bea\label{cKdVa}
\fl  U_{t_{i}t_{j}t_{k}} & = & \frac{1}{2U_{t_{i}}U_{t_{j}}U_{t_{k}}}\bigl(
         U_{t_{i}}U_{t_{j}t_{k}}U_{t_{j}}U_{t_{k}t_{i}}+U_{t_{j}}U_{t_{k}t_{i}}U_{t_{k}}U_{t_{i}t_{j}}
        +U_{t_{k}}U_{t_{i}t_{j}}U_{t_{i}}U_{t_{j}t_{k}}\bigr)\nn\\
\fl   && +\frac{n_{i}^2}{2(t_{k}-t_{i})(t_{i}-t_{j})U_{t_{i}}^2}+\frac{n_{j}^2}{2(t_{i}-t_{j})(t_{j}-t_{k})U_{t_{j}}^2}
         +\frac{n_{k}^2}{2(t_{j}-t_{k})(t_{k}-t_{i})U_{t_{k}}^2}
 \eea
or\footnote{Equations \eqref{cKdVa} and \eqref{cKdVb} are manifestations of the fact that 1+1-dimensional equations of KdV type can be embedded as dimensional reductions of 2+1-dimensional equations of KP type \cite{Vassilis}, which holds true both for the continuous as well as the discrete case. In fact, the continuous nonautonomous equation \eqref{cKdVb} is remarkably similar to the fully discrete Hirota-Miwa equation and we believe that it plays the role of a generating PDE for the KP hierarchy. It would be interesting to see how this equation fits in with the results of \cite{Kono}.}
 \be\label{cKdVb}
  (t_{i}-t_{j})U_{t_{k}}U_{t_{i}t_{j}}+(t_{j}-t_{k})U_{t_{i}}U_{t_{j}t_{k}}+(t_{k}-t_{i})U_{t_{j}}U_{t_{k}t_{i}} = 0.
 \ee
 \ese
Once again the additional equations \eqref{cKdVa} and \eqref{cKdVb} are invoked solely because we are dealing with higher order PDEs in terms of the derivatives, which makes it difficult to verify by direct computation. Nevertheless all three equations \eqref{cKdV}, \eqref{cKdVa} and \eqref{cKdVb} hold on a large class of solutions of soliton type and hence they should certainly be compatible between themselves.

\section{Conclusions}
3-point Lagrangians can be identified for all cases in the ABS list, but in this paper we have restricted ourselves to those cases for which we have established a closure relation. On the basis of this we have formulated a new variational principle in terms of Lagrangian multiforms, which we believe captures the multidimensional consistency of the underlying integrable systems. Obviously we would want to verify that the closure property holds for the remaining cases in the ABS list as well. The cases of Q2 and Q3 have a more implicit structure which makes it more difficult to verify the closure relation directly. Furthermore the case of the top equation Q4 requires the development of new functional identities for the elliptic analogue of the dilogarithm function. We intend to deal with those cases in a separate publication.

An interesting question is whether it is possible to classify integrable discrete and continuous systems on the level of the Lagrangians using the closure property. Furthermore, we envisage that the formalism proposed in this paper would form a paradigm of variational calculus applied to integrable systems, perhaps also in connection with associated physical models. In fact, another interesting question is what is the quantum analogue of this formalism, e.g. in the context of a path integral framework.

\section*{Acknowledgments}
We are grateful to S.N.M. Ruijsenaars for some useful comments. SL was supported by the UK Engineering and Physical Sciences Research Council (EPSRC). This work was completed at the Isaac Newton Institute for Mathematical Sciences, Cambridge, during the programme Discrete Integrable Systems.

\appendix
\section{Dilogarithm functions}
The dilogarithm function is defined by
 \be\label{dilog}
  \Li_2(z) = -\int^z_0{\frac{\ln(1-z)}{z}dz}.
 \ee
Many functional relations involving dilogarithms are given in the book by Lewin\cite{Lewin}, and in the review paper of Kirillov\cite{Kirillov}, which also covers some of the quantum analogues. 
The pivotal functional relation is the five-term identity
 \bea\label{5ptL}
  \Li_2\biggl(\frac{x}{1-y}\frac{y}{1-x}\biggr) & = & \Li_2\biggl(\frac{x}{1-y}\biggr)+\Li_2\biggl(\frac{y}{1-x}\biggr)
                                                     -\Li_2(x)-\Li_2(y)\nn\\
  & & - \ln(1-x)\ln(1-y), \;\;\; x,y<1.
 \eea
For the computations needed for this paper it is more convenient to write \eqref{5ptL} in the form
 \bea\label{5pt}
  \Li_2(s) + \Li_2(t) - \Li_2(st)& = & \Li_2\biggl(\frac{s-st}{1-st}\biggr)+\Li_2\biggl(\frac{t-st}{1-st}\biggr)\nn\\
                                 && +\ln\biggl(\frac{1-s}{1-st}\biggr)\ln\biggl(\frac{1-t}{1-st}\biggr), \;\;\; s,t>1.
 \eea 
An additional two identities needed are the following, both valid for all real $x$.
 \be\label{flip}
   \Li_2(x)+\Li_2\biggl(\frac{1}{x}\biggr) = -\frac{1}{2}\bigl(\ln(-x)\bigr)^2-\frac{\pi^2}{6},
 \ee
 \be\label{2ptL}
  \Li_2(x)+\Li_2\biggl(\frac{x}{x-1}\biggr) = -\frac{1}{2}\bigl(\ln(1-x)\bigr)^2.
 \ee
Equation \eqref{flip} holds regardless of whether the arguments are positive or negative. Equations \eqref{5ptL} and \eqref{2ptL} require additional imaginary terms  depending on the sign of the arguments; these however cancel out in the course of the closure relation calculations.

\section{Proof of the H3 closure relation}
Here we give an outline of the computation needed to show the closure relations \eqref{dclos} hold for H3. We make use of the dilogarithm identities stated in appendix A.
The Lagrangian for H3 is
 \bea
  \cL_{\al_{1}\al_{2}} & \equiv & \cL(u,u_{1},u_{2};\al_{1},\al_{2})\nn\\
                   & = & -\Li_2\biggl(\frac{uu_{1}}{-\al_{1}}\biggr)+\Li_2\biggl(\frac{uu_{2}}{-\al_{2}}\biggr)
                          +\Li_2\biggl(\frac{\al_{2}u_{1}}{\al_{1}u_{2}}\biggr)
                          -\Li_2\biggl(\frac{\al_{1}u_{1}}{\al_{2}u_{2}}\biggr)\nn\\
                      && +\ln\biggl(\frac{\al_{1}^2}{\al_{2}^2}\biggr)\ln(u)+\ln(\al_{2}^2)\ln\biggl(\frac{u_{1}}{u_{2}}\biggr).
 \eea
We make a change of variables, similar to those that appear in the 3 leg form of H3. This will make the computations simpler and easier to follow.
With the abbreviations
 \be
  A = \frac{uu_{1}}{-\al_{1}},\;\;
  B = \frac{uu_{2}}{-\al_{2}},\;\;
  C = \frac{uu_{3}}{-\al_{3}}
 \ee
the Lagrangian becomes
 \bea
  \cL_{\al_{1}\al_{2}} & = & -\Li_2(A)+\Li_2(B)+\Li_2\biggl(\frac{A}{B}\biggr)
                             -\Li_2\biggl(\frac{\al_{1}^2 A}{\al_{2}^2 B}\biggr)\nn\\
                          && +\ln\biggl(\frac{\al_{1}^2}{\al_{2}^2}\biggr)\ln(u)
                             +\ln(\al_{2}^2)\ln\biggl(\frac{u_{1}}{u_{2}}\biggr)
 \eea
whilst the equations of motion, written in the variables $A,B,C$, are as follows:
 \bse
  \bea
   \frac{\al_{1}^2}{\al_{2}^2}\frac{1-A}{1-B} & = & \frac{1-B_{1}}{1-A_{2}}\\
   \frac{\al_{2}^2}{\al_{3}^2}\frac{1-B}{1-C} & = & \frac{1-C_{2}}{1-B_{3}}\\
   \frac{\al_{3}^2}{\al_{1}^2}\frac{1-C}{1-A} & = & \frac{1-A_{3}}{1-C_{1}}.
  \eea
 \ese
Together with the definitions of $A,B,C$, these give expressions for $A_{2},B_{1}$,etc explicitly in terms of $A,B,C$. To write these in a simple way, define the function $H_{A,B}\equiv H(A,B;\al_{1},\al_{2})$ to be
 \be
  H_{A,B} = \frac{\al_{2}^2(1-B)-\al_{1}^2(1-A)}{A-B}
 \ee
leading to the following
 \bea
  && A_{3} = \frac{C}{\al_{1}^2}H_{C,A},\;\; B_{1} = \frac{A}{\al_{2}^2}H_{A,B},\;\; C_{2} = \frac{B}{\al_{3}^2}H_{B,C},\nn\\
  && A_{2} = \frac{B}{\al_{1}^2}H_{A,B},\;\; B_{3} = \frac{C}{\al_{2}^2}H_{B,C},\;\; C_{1} = \frac{A}{\al_{3}^2}H_{C,A}.
 \eea
Defining the quantity $\Gamma$ as below
 \be
  \Gamma \equiv \Dl_{3}\cL_{\al_{1}\al_{2}}+\Dl_{1}\cL_{\al_{2}\al_{3}}+\Dl_{2}\cL_{\al_{3}\al_{1}}
 \ee
we may now write both the Lagrangians and their shifted versions in terms of $A,B$ and $C$, which leads to

\setlength{\unitlength}{1cm}
\begin{picture}(14,12.5)(0.5,6.5)
 \put(0,19){\parbox[t]{16cm}{\bea
\fl \Gamma & = & \;\;\Li_2\biggl(\frac{B}{\al_{1}^2}H_{A,B}\biggr)+\Li_2\biggl(\frac{\al_{1}^2 A}{\al_{2}^2 B}\biggr)
                    -\Li_2\biggl(\frac{A}{\al_{2}^2}H_{A,B}\biggr)\nn\\
\fl           && \nn\\
\fl           && +\Li_2\biggl(\frac{C}{\al_{2}^2}H_{B,C}\biggr)+\Li_2\biggl(\frac{\al_{2}^2 B}{\al_{3}^2 C}\biggr)
                    -\Li_2\biggl(\frac{B}{\al_{3}^2}H_{B,C}\biggr)\nn\\
\fl           && \nn\\
\fl           && +\Li_2\biggl(\frac{A}{\al_{3}^2}H_{C,A}\biggr)+\Li_2\biggl(\frac{\al_{3}^2 C}{\al_{1}^2 A}\biggr)
                    -\Li_2\biggl(\frac{C}{\al_{1}^2}H_{C,A}\biggr)\nn\\
\fl           && \nn\\
\fl           && +\Li_2\biggl(\frac{\al_{1}^2H_{B,C}}{\al_{3}^2H_{A,B}}\biggr)
                    +\Li_2\biggl(\frac{\al_{3}^2H_{A,B}}{\al_{2}^2H_{C,A}}\biggr)
                    +\Li_2\biggl(\frac{\al_{2}^2H_{C,A}}{\al_{1}^2H_{B,C}}\biggr)\nn\\
\fl           && \nn\\
\fl           && -\Li_2\biggl(\frac{H_{B,C}}{H_{A,B}}\biggr)-\Li_2\biggl(\frac{H_{A,B}}{H_{C,A}}\biggr)
                    -\Li_2\biggl(\frac{H_{C,A}}{H_{B,C}}\biggr)\nn\\
\fl           && \nn\\
\fl           && -\Li_2\biggl(\frac{A}{B}\biggr)-\Li_2\biggl(\frac{B}{C}\biggr)-\Li_2\biggl(\frac{C}{A}\biggr)\nn\\
\fl           && \nn\\
\fl           && +\ln\biggl(\frac{\al_{3}^2}{\al_{1}^2}\biggr)\ln(H_{A,B})
                    +\ln\biggl(\frac{\al_{1}^2}{\al_{2}^2}\biggr)\ln(H_{B,C})
                    +\ln\biggl(\frac{\al_{2}^2}{\al_{3}^2}\biggr)\ln(H_{C,A})\nn\\
\fl           && -\ln\biggl(\frac{\al_{3}^2}{\al_{1}^2}\biggr)\ln(A)-\ln\biggl(\frac{\al_{1}^2}{\al_{2}^2}\biggr)\ln(B)
                    -\ln\biggl(\frac{\al_{2}^2}{\al_{3}^2}\biggr)\ln(C)\nn\\
\fl           && -\ln(\al_{1}^2)\ln(\al_{2}^2)-\ln(\al_{2}^2)\ln(\al_{3}^2)-\ln(\al_{3}^2)\ln(\al_{1}^2)\nn\\
\fl           && +(\ln(\al_{1}^2))^2+(\ln(\al_{2}^2))^2+(\ln(\al_{3}^2))^2\nn
 \eea}}
 \put(0.7,17.5){\framebox(8.8,1.3)}
 \put(0.7,16.1){\framebox(8.8,1.2)}
 \put(0.7,14.7){\framebox(8.8,1.2)}
 \put(0.7,13.2){\framebox(9.4,1.3)}
 \put(0.7,11.8){\framebox(8.2,1.3)}
 \put(0.7,10.4){\framebox(6.5,1.3)}
 \put(6.9,13.3){\dashbox{0.1}(3.1,1.1)}
 \put(6.1,11.9){\dashbox{0.1}(2.7,1.1)}
 \put(5,10.5){\dashbox{0.1}(2.1,1.1)}
\end{picture}\\
where we have we rearranged the terms in a way that suggests which dilogarithm identities to use and where.
Applying the dilogarithm identity \eqref{flip} to the terms in the dashed-line boxes, the argument of the dilogarithm functions can be inverted. This enables us to use identity \eqref{5pt} on the terms grouped in the solid-line boxes, using the definition of $H_{A,B}$ to simplify the outcome. We will gather all the logarithm terms together at the end.\\
\setlength{\unitlength}{1cm}
\begin{picture}(14,6)(0,0)
 \put(0,6){\parbox[t]{6cm}{\bea
\fl  \Gamma & = & +\Li_2\biggl(\frac{(A-B)H_{A,B}}{\al_{1}^2(A-1)}\biggr)\;\;
               +\Li_2\biggl(\frac{A(B-1)}{B(A-1)}\biggr)\nn\\
\fl         && +\Li_2\biggl(\frac{(B-C)H_{B,C}}{\al_{2}^2(B-1)}\biggr)\;\;+\Li_2\biggl(\frac{B(C-1)}{C(B-1)}\biggr)\nn\\
\fl         && +\Li_2\biggl(\frac{(C-A)H_{C,A}}{\al_{3}^2(C-1)}\biggr)\;\;+\Li_2\biggl(\frac{C(A-1)}{A(C-1)}\biggr)\nn\\
\fl         && \nn\\
\fl         && +\Li_2\biggl(\frac{\al_{1}^2(A-1)(B-C)H_{B,C}}{\al_{3}^2(C-1)(B-A)H_{A,B}}\biggr)\;
               +\Li_2\biggl(\frac{(C-A)(B-1)}{(B-A)(C-1)}\biggr)\nn\\
\fl         && -\Li_2\biggl(\frac{(B-C)H_{B,C}}{(B-A)H_{A,B}}\biggr)-\Li_2\biggl(\frac{C-A}{B-A}\biggr)\nn\\
\fl         && -\Li_2\biggl(\frac{A(B-C)}{B(A-C)}\biggr)-\Li_2\biggl(\frac{A-B}{A-C}\biggr)\nn
 \eea}}
 \put(0.8,4.6){\framebox(4.1,1.2)}
 \put(5.1,2.2){\framebox(3.7,3.6)}
 \put(0.8,0.8){\framebox(5.9,1.2)}
 \put(5.2,2.3){\dashbox{0.1}(3.4,1.1)}
 \put(6.9,0.8){\dashbox{0.1}(4.6,1.2)}
 \put(4.4,-1.5){\dashbox{0.1}(3,1.2)}
 \put(0.8,3.3){\oval(1.2,3)[l]}
\end{picture}

\bea
\fl         && +\ln\biggl(\frac{\al_{2}^2(B-1)}{\al_{1}^2(A-1)}\biggr)\ln\biggl(\frac{A-B}{B(A-1)}\biggr)
               +\ln\biggl(\frac{\al_{3}^2(C-1)}{\al_{2}^2(B-1)}\biggr)\ln\biggl(\frac{B-C}{C(B-1)}\biggr)\nn\\
\fl         && +\ln\biggl(\frac{\al_{1}^2(A-1)}{\al_{3}^2(C-1)}\biggr)\ln\biggl(\frac{C-A}{A(C-1)}\biggr)\nn\\
\fl         && +\ln\biggl(\frac{\al_{2}^2(B-1)(C-A)H_{C,A}}{\al_{3}^2(C-1)(B-A)H_{A,B}}\biggr)
                \ln\biggl(\frac{(A-1)(B-C)}{(C-1)(B-A)}\biggr)\nn\\
\fl         && -\ln\biggl(\frac{(C-A)H_{C,A}}{(B-A)H_{A,B}}\biggr)\ln\biggl(\frac{B-C}{B-A}\biggr)
               -\ln\biggl(\frac{C(A-B)}{B(A-C)}\biggr)\ln\biggl(\frac{B-C}{A-C}\biggr)\nn\\
\fl         && -\frac{1}{2}\biggl(\ln\biggl(-\frac{\al_{1}^2H_{B,C}}{\al_{2}^2H_{C,A}}\biggr)\biggr)^2
               +\frac{1}{2}\biggl(\ln\biggl(-\frac{H_{B,C}}{H_{C,A}}\biggr)\biggr)^2
               +\frac{1}{2}\biggl(\ln\biggl(-\frac{A}{C}\biggr)\biggr)^2\nn\\
\fl         && +\frac{\pi^2}{6}+\ln\biggl(\frac{\al_{3}^2}{\al_{1}^2}\biggr)\ln(H_{A,B})
               +\ln\biggl(\frac{\al_{1}^2}{\al_{2}^2}\biggr)\ln(H_{B,C})
               +\ln\biggl(\frac{\al_{2}^2}{\al_{3}^2}\biggr)\ln(H_{C,A})\nn\\
\fl         && -\ln\biggl(\frac{\al_{3}^2}{\al_{1}^2}\biggr)\ln(A)-\ln\biggl(\frac{\al_{1}^2}{\al_{2}^2}\biggr)\ln(B)
               -\ln\biggl(\frac{\al_{2}^2}{\al_{3}^2}\biggr)\ln(C)\nn\\
\fl         && -\ln(\al_{1}^2)\ln(\al_{2}^2)-\ln(\al_{2}^2)\ln(\al_{3}^2)-\ln(\al_{3}^2)\ln(\al_{1}^2)\nn\\
\fl         && +(\ln(\al_{1}^2))^2+(\ln(\al_{2}^2))^2+(\ln(\al_{3}^2))^2.
\eea
Again, using identity \eqref{flip} on the terms in the dashed-line boxes, and subsequently identity \eqref{5pt} on the terms grouped in the solid-line boxes, we obtain
 \bea
\fl  \Gamma & = & \;\;\Li_2\biggl(\frac{(C-B)H_{B,C}}{\al_{3}^2(C-1)}\biggr)
               +\Li_2\biggl(\frac{(A-C)H_{C,A}}{\al_{1}^2(A-1)}\biggr)\nn\\
\fl         && +\Li_2\biggl(\frac{(B-C)H_{B,C}}{\al_{2}^2(B-1)}\biggr)
               +\Li_2\biggl(\frac{(C-A)H_{C,A}}{\al_{3}^2(C-1)}\biggr)\nn\\
\fl         && +\ln\biggl(\frac{\al_{2}^2(B-1)}{\al_{1}^2(A-1)}\biggr)\ln\biggl(\frac{A-B}{B(A-1)}\biggr)
               +\ln\biggl(\frac{\al_{3}^2(C-1)}{\al_{2}^2(B-1)}\biggr)\ln\biggl(\frac{B-C}{C(B-1)}\biggr)\nn\\
\fl         && +\ln\biggl(\frac{\al_{1}^2(A-1)}{\al_{3}^2(C-1)}\biggr)\ln\biggl(\frac{C-A}{A(C-1)}\biggr)\nn\\
\fl         && +\ln\biggl(\frac{\al_{2}^2(B-1)(C-A)H_{C,A}}{\al_{3}^2(C-1)(B-A)H_{A,B}}\biggr)
                \ln\biggl(\frac{(A-1)(B-C)}{(C-1)(B-A)}\biggr)\nn\\
\fl         && -\ln\biggl(\frac{(C-A)H_{C,A}}{(B-A)H_{A,B}}\biggr)\ln\biggl(\frac{B-C}{B-A}\biggr)
               -\ln\biggl(\frac{C(A-B)}{B(A-C)}\biggr)\ln\biggl(\frac{B-C}{A-C}\biggr)\nn\\
\fl         && +\ln\biggl(\frac{C(A-B)}{B(A-C)}\biggr)\ln\biggl(\frac{(A-1)(B-C)}{(B-1)(A-C)}\biggr)
               +\ln\biggl(\frac{\al_{3}^2(C-1)}{\al_{1}^2(A-1)}\biggr)
                \ln\biggl(\frac{(C-A)H_{C,A}}{(B-A)H_{A,B}}\biggr)\nn\\
\fl         && -\frac{1}{2}\biggl(\ln\biggl(-\frac{\al_{1}^2H_{B,C}}{\al_{2}^2H_{C,A}}\biggr)\biggr)^2
               +\frac{1}{2}\biggl(\ln\biggl(-\frac{H_{B,C}}{H_{C,A}}\biggr)\biggr)^2
               +\frac{1}{2}\biggl(\ln\biggl(-\frac{A}{C}\biggr)\biggr)^2\nn\\
\fl         && -\frac{1}{2}\biggl(\ln\biggl(-\frac{A(C-1)}{C(A-1)}\biggr)\biggr)^2
               -\frac{1}{2}\biggl(\ln\biggl(-\frac{(B-A)(C-1)}{(C-A)(B-1)}\biggr)\biggr)^2
               +\frac{1}{2}\biggl(\ln\biggl(-\frac{A-C}{A-B}\biggr)\biggr)^2\nn\\
\fl         && +\ln\biggl(\frac{\al_{3}^2}{\al_{1}^2}\biggr)\ln(H_{A,B})
               +\ln\biggl(\frac{\al_{1}^2}{\al_{2}^2}\biggr)\ln(H_{B,C})
               +\ln\biggl(\frac{\al_{2}^2}{\al_{3}^2}\biggr)\ln(H_{C,A})\nn
 \eea
 \bea\label{gm9}
\fl         && -\ln\biggl(\frac{\al_{3}^2}{\al_{1}^2}\biggr)\ln(A)-\ln\biggl(\frac{\al_{1}^2}{\al_{2}^2}\biggr)\ln(B)
               -\ln\biggl(\frac{\al_{2}^2}{\al_{3}^2}\biggr)\ln(C)\nn\\
\fl         && -\ln(\al_{1}^2)\ln(\al_{2}^2)-\ln(\al_{2}^2)\ln(\al_{3}^2)-\ln(\al_{3}^2)\ln(\al_{1}^2)\nn\\
\fl         && +(\ln(\al_{1}^2))^2+(\ln(\al_{2}^2))^2+(\ln(\al_{3}^2))^2.
 \eea
Using identity \eqref{2ptL} on the first term of line 1 and the second term of line 2 of \eqref{gm9} leaves the dilogarithm terms which subsequently cancel out. What then remains are only the logarithm terms, which also cancel out, leaving $\Gamma=0$. This concludes the proof of the closure relation.

\end{document}